\documentclass[12pt,a4paper]{article}
\usepackage{a4wide}
\usepackage{latexsym}
\usepackage{epsf}
\usepackage{amssymb}
\begin{document}
\def\be{\begin{equation}}
\def\ee{\end{equation}}
\def\bea{\begin{eqnarray}}
\def\eea{\end{eqnarray}}

\def\pd{\partial}
\def\a{\alpha}
\def\b{\beta}
\def\g{\gamma}
\def\d{\delta}
\def\m{\mu}
\def\n{\nu}
\newcommand{\fsl}{{\hspace{-7pt}\slash}}
\newcommand{\gsl}{{\hspace{-5pt}\slash}}
\newcommand{\dslash}{\pd\fsl}
\newcommand{\pslash}{p\gsl }
\newcommand{\Aslash}{A\gsl }
\newcommand{\qslash}{q\gsl }
\newcommand{\Lslash}{\Lambda\gsl }
\newcommand{\kuslash}{k_1\gsl }
\newcommand{\kdslash}{k_2\gsl }
\newcommand{\Dslash}{D\fsl}
\def \h{\mathcal{H}}
\def \hh{\mathcal{G}}
\def\t{\tau}
\def\p{\pi}
\def\th{\theta}
\def\l{\lambda}
\def\O{\Omega}
\def\r{\rho}
\def\s{\sigma}
\def\e{\epsilon}
  \def\scri{\mathcal{J}}
\def\cM{\mathcal{M}}
\def\tcM{\tilde{\mathcal{M}}}
\def\RR{\mathbb{R}}

\hyphenation{re-pa-ra-me-tri-za-tion}
\hyphenation{trans-for-ma-tions}


\begin{flushright}
IFT-UAM/CSIC-06-06\\
hep-th/0602150\\
\end{flushright}

\vspace{1cm}

\begin{center}

{\bf\Large  Renormalized Kaluza-Klein theories.}

\vspace{.5cm}

{\bf Enrique \'Alvarez and Ant\'on F. Faedo }

\vspace{.3cm}

\vskip 0.4cm  
 
{\it  Instituto de F\'{\i}sica Te\'orica UAM/CSIC, C-XVI,
and  Departamento de F\'{\i}sica Te\'orica, C-XI,\\
  Universidad Aut\'onoma de Madrid 
  E-28049-Madrid, Spain }

\vskip 0.2cm

\vskip 1cm

{\bf Abstract}
\par

Using six-dimensional quantum electrodynamics ($QED_6$) as an example we study the one-loop renormalization of the
theory both from the six and four-dimensional points of view. Our main conclusion is that the
properly renormalized four dimensional theory never forgets its higher dimensional origin. In 
particular, the coefficients of the neccessary 
extra counterterms in the four dimensional theory are determined in a precise way. We check our results by studying
the reduction of $QED_4$ on a two-torus.
\end{center}

\begin{quote}

\end{quote}


\newpage

\setcounter{page}{1}
\setcounter{footnote}{0}
\tableofcontents
\newpage
\section{Introduction}
The origin of four dimensional gauge symmetries is one of the deepest mysteries of physics.
The idea of Theodor Kaluza, improved by Oskar Klein (cf. for example, \cite{Chodos} and references therein) 
that higher dimensional
spacetime symmetries imply low energy gauge symmetries in four dimensions provided the extra
dimensions are curled up in an appropiate way has proved quite fruitful and worth pursuing.
\par
In the simplest setting, the Einstein-Hilbert gravitational 
action in a five-dimensional manifold
which is a product of four-dimensional Minkowski space-time with a one-dimensional circle
of radius $R$, looks at energies
\be
E<< M\equiv \frac{1}{R}
\ee
like four-dimensional Einstein-Hilbert coupled to an abelian Maxwell field. 
\par
In order to be more precise, if we believe that extra dimensions are {\em real}, we got to 
renormalize the theory. Even if we do not embed the extra-dimensional theory in some suposedly 
consistent framework, such as supestrings, (which would provide a cutoff of sorts), at one loop order, the fact
 that the higher dimensional (sometimes called the mother) theory is not renormalizable 
is not directly relevant, in 
the sense
that we still can study and classify all divergences. For example, the six-dimensional 
electric charge is 
 dimensionful, which allows  for an unbounded number 
of candidate counterterms. However, to any given order in perturbation theory 
this number is finite, and the theory can in principle be renormalized, although it is
 still true that always appear new operators in the counterterms which were not
present in the original lagrangian. This is then essentially a low energy
approximation, because we can only expect it to be good (in the example of $QED_6$, in which
we are going to concentrate upon) when the dimensionless
quantity $\a_{d=6}E^2 << 1$, where $\a_{d=6}$ is the six-dimensional fine structure
constant. Given the fact that the six and four-dimensional coupling constants are related by
$\a_{d=6}M^2\equiv\a_{d=4}\equiv\frac{1}{137}$, in terms of the usual four-dimensional fine-structure constant, this means
$E<<\frac{M}{\sqrt{\a}}\sim 10 M$. It follows that one can compute reliably for energies
$E\sim M$, but not much bigger. Our viewpoint will thus be that the
theory is {\em defined} in higher dimensions by means of the necessary counterterms, in a sense
that we shall try to make more precise in what follows.
\par
At any rate, and in order to dissipate any doubts, we shall repeat in due course the same analysis on $QED_4$ on a two-torus.
In this case the extra dimensional theory is well defined (forgetting the Landau pole), and our results are
essentially the same.
\par
Besides the six-dimensional viewpoint we are going to favor, there is always the possibility of
expanding all fields in harmonics and perform the integrals over the extra dimensional compact manifold.
In that way we find a four dimensional theory, but with an infinite number of fields. It seems quite intuitive that
provided we keep track of the infinite set of modes, this four dimensional theory should be equivalent to
the full extra-dimensional one; their respective divergences, in particular, should match.
The main purpose of this paper is to check this intuition with some explicit computations. 
Although we are not going to work  it out in any detail, it should be possible to express 
our results in 
the language of effective low energy field theories. Some steps in this direction have
been already given in \cite{Oliver} \cite{Ghilencea}.
\par
There are then two complementary
viewpoints, the higher dimensional one, and the four dimensional with the Kaluza-Klein
 tower, and if we want to make explicit statements on exactly when the tower begins to be 
relevant, we have to relate not only the classical parts, but also the quantum contributions
 on both sides. 
\par
Curiously enough, in the case the fields only interact through the universal coupling to an external gravitational field,
the two viewpoints are {\em exactly} equivalent (with some qualifications).This was proved by Duff and Toms \cite{Duff}, and provided
a strong motivation for our research.
\par
We shall work to one loop order only. To this order,  the effective action is
 given in terms of a functional determinant. We shall regularize it through the heat kernel
approach, which respects all gauge invariances, including the geometrical ones. Let us quickly
review our notation and remark on some potential ambiguities.
\par
The geometric setting is given by a riemannian n-dimensional manifold, with a metric
$g_{MN}$.  
This manifold will usually be of a factorized form: $\mathbb{R}^4\times K$
where $K$ is a compact $(n-4)$-dimensional manifold , and $\mathbb{R}^4$ 
represents the euclidean version of Minkowski space.
 More generally, like in the models recently popularized by
Randall and Sundrum \cite{Randall}, this structure is present only locally, 
i.e., we have a fiber bundle 
(warped space) based on
Minkowski space.
\par

All our operators enjoy the form
\be\label{ope}
\Delta\equiv -D_{M}D^{M} + Y
\ee
with 
\be
D_M\equiv \pd_M+X_M
\ee
and the operator defining the heat kernel is formally given by:
\be
K(\t)\equiv e^{-\t \Delta}
\ee
acting on a convenient functional space in such a way that
\be
(K f)(x)\equiv \int \sqrt{|g|} d^n y K(x,y;\t) f(y)
\ee
The short time off-diagonal  \cite{de Witt} expansion is 
defined (for manifolds without
 boundary) by:
\be
K(x,y;\t)=K_0(x,y;\t)\sum_{p=0}b_{2p}(x,y) \t^p
\ee
where the flat space solution is given by:
\be
K_0(x,y;\t)=\frac{1}{(4\pi\t)^{n/2}}e^{-\frac{\s^2}{4\t}}
\ee
and $\s$ is the geodesic distance
between the two points, given in flat space by:
\be
\s^2=(x-y)^2
\ee
and for consistency
\be
b_0(x,x)=1.
\ee
When boundaries are present, odd powers of $\t^{1/2}$ do appear, which can 
formally be incorporated in the
former expansion by allowing non vanishing odd coefficients, $b_{2p+1}\t^{p+ 1/2}\neq0$.
\par
It is sometimes useful to consider the integrated quantity:
\be
Y(\t,f)\equiv tr\,(f e^{-\t \Delta})=\sum_{k=0}\t^{\frac{k-n}{2}}a_k(f)
\ee
where the trace involves whatever finite rank indices the operator might posses, and 
\be
a_k(f)=\frac{1}{(4\pi)^{n/2}}\int \sqrt{|g|}d^n x\,tr\,b_k(x,x)f(x)
\ee
The mass dimension of $a_k$ is $k-n$, whereas the one of $b_k$ is simply $k$.
It follows that
\be
a_0=\frac{tr\,{\mathbb{I}}}{(4\pi)^{n/2}}V\equiv\frac{1}{(4\pi)^{n/2}}\int \sqrt{|g|}d^n x\,tr\mathbb{I}
\ee
As usual, we shall denote
\be
a_k\equiv a_k(f=1)
\ee
Note in particular that
\be
Y(\t)\equiv Y(\t,f=1)=tr\, e^{-\t \Delta}=\sum_{k=0}\t^{\frac{k-n}{2}}a_k
\ee
After all these prolegomena,
the determinant is defined as:
\be
\log\,\det\,\Delta=-\int_0^\infty\frac{d\t}{\t^{1+ n/2}}\sum_{p=0}a_{p}\tau^{p/2}
\ee
There are several possible viewpoints on this integral. One of them is to analytically continue
on the dimension  $n$.
The integral over the proper time $\t$, cut off in the infrared by $\t_{max}=\m^{-2}$ 
produces poles in 
the complex variable $n$, given by:
\be
\log\,\det\,\Delta=-  \sum_{p=0}a_{p}\frac{2 \m^{n-p}}{p-n}+finite\,part.
\ee
which when $n$ approaches the physical dimension, say, $d$,
\be
n=d+\e
\ee
yields the divergent piece of the determinant (a dimensionless quantity):
\be
\log\,\det\,\Delta |_{div}=\frac{2 \m^\e}{ \e}\,a_d(\Delta).
\ee
in the even dimensional case. This prescription yields
 a finite answer for  odd dimensions in the absence of a boundary, and is the one usually favored when
 working with effective lagrangians (cf. for example \cite{Donoghue}).
\par
A different, and in some sense more physical possibility is to 
introduce a cutoff in the lower end of the proper time integral,
 $\Lambda/\m\rightarrow\infty$.
 In that way we get, for example \footnote{Although we shall try our best 
to avoid cluttering the notation unnecessarily,
 we are forced to distinguish between quantities bearing identical names, 
but coming from different dimensions.} in six dimensions:
\be\label{6d}
\log\,\det\,\Delta |_{div}=\frac{1}{3} a_0\Lambda_{(d=6)}^6+ \frac{1}{2}
 a_2 \Lambda_{(d=6)}^4+  a_4\Lambda_{(d=6)}^2+ a_6 \log{\frac{\Lambda^2_{(d=6)}}{\m^2_{(d=6)}}}
\ee
Where the heat kernel coefficients are obviously in six dimensions. In four dimensions instead:
\be\label{4d}
\log\,\det\,\Delta |_{div}=\frac{1}{2} a_0\Lambda_{(d=4)}^4+  a_2 \Lambda_{(d=4)}^2+
  a_4 \log{\frac{\Lambda^2_{(d=4)}}{\m^2_{(d=4)}}}
\ee
Where now the coefficients are the corresponding ones in four dimensions. The dominant divergence (sixth power and fourth power of the cutoff) is universal and 
independent of the particular operator under consideration. We shall not study it
further here.

\par
In spite of the fact that it is often pointed out that there is no way of
imposing a cutoff in a gauge invariant way, we would like to stress that, at least to the
one loop order, this procedure respects all gauge invariances, abelian and non
abelian, as well as general covariance in its case. This is obvious, because we are {\em not} cutting off the momentum,
but rather the {\em proper time}, a covariant as well as gauge invariant concept. If we remember that the proper time in the sense we are employing it,
has mass dimension $-2$, we are neglecting in the evaluation of the one loop determinants proper times smaller than $\Lambda^{-1}$. This fact, which was probably first pointed out by Schwinger
\cite{Schwinger} in 1951, has been exploited by Bryce deWitt \cite{de Witt} to get covariant expansions in quantum gravity; and also by
Fujikawa \cite{Fujikawa} to get the covariant anomaly.
\par 
We shall denote these two procedures {\em dimensional regularization} and {\em cutoff}, 
respectively. Both respect all gauge invariances of the theory but only the cutoff theory
yields information on the divergences in the odd dimensional case. 
\section{Six dimensional quantum electrodynamics compactified on a torus.}

 Let us now consider an example not altogether trivial, namely quantum electrodynamics (QED) on 
a six-dimensional 
manifold which is 
topologically four-dimensional 
Minkowski space times a two-torus, that is, $\mathbb{R}^4\times S^1\times S^1$. 
This example avoids the complications of interacting gravitational sectors, but in some 
sense is not representative of
the whole Kaluza-Klein philosophy, because we are introducing gauge fields already in the extra dimensions. We are using it as a toy model.
\par
The metric for the
time being is assumed to be
\be
ds^2=\delta_{\m\n}dx^{\m}dx^\n+R_5^2 d\theta_5^2 + R_6^2 d\theta_6^2
\ee
that is, $y_5=R_5 \theta_5$ and  $y_6=R_6 \theta_6$. We shall follow consistently the 
convention that capital indices, like $M,N,\ldots$ run over the full dimensions, in our case 
from $1$ to $6$; greek indices, $\m,\n,\ldots$ run over the ordinary Minkowski coordinates, from $1$ to $4$; and small roman letters, 
$a,b,\ldots$, over the extra dimensions, that is, from $5$ to $6$.
\par
The (euclidean version of the) action then reads
\be
S=\int d^6 x\left(\frac{1}{4}F_{MN}^2+\bar{\psi}(\Dslash+m)\psi\right)
\ee
where the abelian covariant derivative is simply:
\be
D_M\psi\equiv\left(\pd_M-eA_M\right)\psi
\ee
Let us recall here that, for vanishing curvature, the general formulas \cite{Gilkey} 
for the first few
 coefficients of an operator of the form (\ref{ope}) are:
\be
a_2=-\int\frac{d^n x}{(4\pi)^{\frac{n}{2}}}\,tr\, Y
\ee
\be
a_4=\int\frac{d^n x}{(4\pi)^{\frac{n}{2}}}\,tr\left(\frac{1}{12}X_{MN}^2+\frac{1}{2}Y^2-
\frac{1}{6}Y_{;MM}\right)
\ee
\begin{eqnarray}\label{eq:a6}
\lefteqn{a_6=\frac{1}{360}\int\frac{d^n x}{(4\pi)^{\frac{n}{2}}}
\,tr\Big( 8X_{MN;R}^2+2X_{MN;N}^2+{} } \nonumber\\& & {}+12X_{MN;RR}X^{MN}-12X_{MN}X^{NR}
X_R^{\hspace{1ex}M}-6Y_{;MMNN}+{} \nonumber\\& & {}+60YY_{;MM}+30Y_{;M}^2-
60Y^3-30YX_{MN}^2\Big)
\end{eqnarray}
Where $;$ denotes covariant derivative, and
\be
X_{MN}=\pd_M X_N-\pd_N X_M +\left[X_M,X_N\right]
\ee
In order to perform the explicit computation, it is exceedingly useful to combine the fermionic
and bosonic sectors in a full supermatrix. Please read the appendix for a brief review of
the  technique and notation.
\par

Computing the coefficients is then straightforward albeit somewhat laborious.
In terms of the background fields $\bar{A}_M,\eta,\bar{\eta}$
\be
a_2=\int\frac{d^6 x}{(4\pi)^3}8m^2
\ee
as well as
\be
 a_4=\int\frac{d^6 x}{(4\pi)^3}\left(\frac{4}{3}e^2\bar{F}_{MN}^2+4 e^2 \bar{\eta}\bar{\Dslash}\eta
+12me^2\bar{\eta}\eta\right)
\ee
 Finally  we get (using  the background equations of motion):
\begin{eqnarray}\label{sdcounter}
\lefteqn{ a_6=\int\frac{d^6 x}{(4\pi)^3}\left(-\frac{1}{12}e^4\bar{\eta}\Sigma_{MNL}\eta\bar{\eta}\Sigma^{MNL}\eta+\frac{19}{15}e^2m\bar{\eta}\bar{D}_M\bar{D}^M\eta+\frac{2}{15}e^3\bar{\eta}\gamma_N\bar{D}_M\eta\bar{F}^{MN}-\right.}\nonumber\\
& &{}-e^3m\bar{\eta}\gamma_M\gamma_N\eta\bar{F}_{MN}-2e^2m^2\bar{\eta}
\gamma^M\bar{D}_M\eta-6e^2m^3\bar{\eta}\eta-\frac{11}{45}e^2\bar{D}_R\bar{F}_{MN}\bar{D}^R\bar{F}^{MN}+{}\nonumber\\
&&\left.{}+\frac{23}{9}e^2\bar{D}_M\bar{F}^{MN}\bar{D}^R\bar{F}_{RN}-\frac{4}{3}
e^2m^2\bar{F}_{MN}\bar{F}^{MN}\right)
\end{eqnarray}
Where $\Sigma_{MNL}$ is the totally antisymmmetric product of three gammas. Remember that in dimensional regularization 
\be
\Delta S=\frac{1}{\e}a_6
\ee
plus a possible finite part. With a cutoff, these are the logarithmic divergences, and
we have in addition both quadratic and quartic divergences, on which more to follow.
\par
The first conclusion we can draw from this analysis is that
 quantum effects, besides renormalizing the six-dimensional couplings, induce a set of 
non-minimal interactions which are generated with arbitrary coefficients.
\par
Actually, due to the fact that the mass dimension of the coupling constant is $-1$, there is 
no  finite closed set of operators of counterterms. Let us be more specific.
\par
First of all, there is a dimension five operator, which becomes a potential
counterterm  in the massive case:
\be
{\cal O}_{(5)}=\left(\bar{\psi}\psi\right)
\ee
The set of gauge-invariant dimension six operators is given by: 
\be
{\cal O}^i_{(6)}=\left(\bar{\psi} \Dslash\psi,\,F_{MN}^2\right)
\ee
To the next order, that is, dimension seven, the list reads: 
\be
{\cal O}^i_{(7)}=\left(\bar{\psi} \Dslash\Dslash\psi\right)
\ee
The dimension eight operators are: 
\be
{\cal O}^i_{(8)}=\left(\bar{\psi} \Dslash\Dslash\Dslash\psi,\,
\bar{\psi}\s_{MN}\psi F^{MN},\,D^M F_{MN}D_RF^{RN},\,F_{NL}D^2F^{NL}\right)
\ee
And finally, to dimension nine we have to consider:
\be
{\cal O}^i_{(9)}=\left(\bar{\psi} \g_M D_N\psi F^{MN},\,\bar{\psi}D_A D_B D_C D_D\psi t^{ABCD}
\right)
\ee

In the massive case the dimension of this operators can be increased by introducing powers of $m$. Amongst the operators that actually appear as counterterms only the ${\cal O}^2_{(8)}$ is absent. At any rate it should be plain
that we can claim results only to first nontrivial order in the six-dimensional fine structure constant, and
that we have really no right to keep the $e^3$ and $e^4$ terms in the counterterm.
\par
The non renormalizabity of the theory manifests itself in the fact that if we were to include
all those dimension seven and dimension eight operators, they would generate more and more
higher dimension operators as counterterms. There is no closed set, unless we assume, as is
natural to the order we are working, that the effect of all those couplings is of higher
order in the six-dimensional fine structure constant.\footnote{
Keeping in mind that we are not performing a fully consistent computation,
if we define the renormalization constants as is usually done:
\bea
&&A_0=Z_3^{1/2}A\nonumber\\
&&\psi_0=Z_2^{1/2}\psi\nonumber\\
&&e_0=Z_1 Z_2^{-1} Z_3^{-1/2} e\nonumber\\
&&m_0=Z_m m
\eea
we easily get $Z_1=Z_2$ which conveys the fact that the theory is gauge invariant, and
\bea
&&Z_2=1-\frac{ e^2 m^2}{32\pi^3\e}\nonumber\\
&&Z_3=1-\frac{ e^2 m^2}{12\pi^3 \e}\nonumber\\
&&Z_m=1-\frac{ e^2 m^2}{16\pi^3\e}
\eea

A simple calculation then leads to the renormalization group functions:
\bea
&&\b_e\equiv\frac{\pd e}{\pd\log{\m}}=-\frac{1}{24 \pi^3} e^3 m^2\nonumber\\
&&\b_m\equiv\frac{\pd m}{\pd \log{\m}}=\frac{1}{16\pi^3} e^2 m^3
\eea
The renormalization of the fermion mass is entangled with the charge renormalization.
The behavior of the coupling constants reads:
\bea
&&e=e_0-\frac{1}{24\pi^3} m_0^2 e_0^3 \log{\m/\m_0}\nonumber\\
&&m=m_0\left(1-\frac{1}{24 \pi^3} m_0^2 e_0^2 \log{\m/\m_0}\right)^{- 3/2}
\eea
The dimensionful charge vanishes when
\be
\m=\m_0 e^{\frac{24\pi^3}{m_0^2 e_0^2}}
\ee
If we define the dimensionless couplings
\bea
&&\hat{e}\equiv e\m\nonumber\\
&&\hat{m}\equiv\frac{m}{\m}
\eea
then the renormalization group equations read
\bea
&&\b_{\hat{e}}=\hat{e}-\frac{1}{24\pi^3}\hat{m}^2 \hat{e}^3\nonumber\\
&&\b_{\hat{m}}=-\hat{m}+\frac{1}{16\pi^3}\hat{e}^2\hat{m}^3
\eea
}

\section{The four-dimensional viewpoint}

Let's consider the point of view of the reduced theory. We can now expand all fields in
 Fourier series:
\be
\phi (x,y)=\frac{1}{2\pi\sqrt{R_5 R_6}}\sum_n \phi_n(x) e^{i \frac{n}{R}.y}
\ee
where $n\equiv (n_5,n_6)$, and we have included a convenient factor in front to
take care of the diference of canonical dimensions of the fields in six and four dimensions.
Real fields (such as the photon) obey
\be
\phi^*_n(x)=\phi_{-n}(x)
\ee
The six-dimensional gamma matrices can be chosen such as:
\bea
&&\g_\m^{(6)}=\s_3\otimes\g_\m^{(4)}\nonumber\\
&&\g_5^{(6)}=\s_1\otimes 1\nonumber\\
&&\g_6^{(6)}=\s_2\otimes1
\eea
In that way, six-dimensional spinors split in two four-dimensional ones:
\be
\psi=\left(\begin{array}{c}
\psi_1\\
\psi_2
\end{array}\right)
\ee
It is a simple matter to perform the integrals over the angular variables and obtain
the gauge fixed action (still exact) in the four dimensional form:
\bea\label{tower}
\lefteqn{S=\int d^4 x\sum_{n_5,n_6}\left(\bar{\psi}^1_{n}\dslash \psi^1_n+\bar{\psi}^2_{n}
\dslash \psi^2_n+\bar{\psi}^1_{n}(i\frac{n_5}{R_5}+\frac{n_6}{R_6})\psi^2_n-
\bar{\psi}^2_{n}(i\frac{n_5}{R_5}-\frac{n_6}{R_6})\psi^1_n+\right.{}}\nonumber\\
&&{}+m\left(\bar{\psi}^1_{n}\psi^1_n-\bar{\psi}^2_{n}\psi^2_n\right)
-\frac{1}{2}(A_\m^n)^{*}\left(\Box-\frac{n_5^2}{R_5^2}-\frac{n_6^2}{R_6^2}\right) A^\m_n-
\frac{1}{2}(A_5^n)^{*}\left(\Box-\frac{n_5^2}{R_5^2}-\frac{n_6^2}{R_6^2}\right)A_5^n-
\nonumber\\
&&\left.-\frac{1}{2}(A_6^n)^{*}\left(\Box-\frac{n_5^2}{R_5^2}-
\frac{n_6^2}{R_6^2}\right)A_6^n-e\sum_m\left(\bar{\psi}^1_m\Aslash_{m-n}\psi^1_n+
\bar{\psi}^2_m\Aslash_{m-n}\psi^2_n
+\right.\right.{}\nonumber\\&&{}\left.+\bar{\psi}^1_mA_5^{m-n}\psi^2_n-
\bar{\psi}^2_mA_5^{m-n}\psi^1_n-i\bar{\psi}^1_mA_6^{m-n}\psi^2_n-
i\bar{\psi}^2_mA_6^{m-n}\psi^1_n\right)\Bigg)
\eea
and the four-dimensional coupling constant is
\be
e\equiv \frac{e^{(6)}}{2\pi\sqrt{R_5 R_6}}\equiv e^{(6)}M
\ee
Here we see clearly a generic feature of interacting theories, namely that there is no
consistent truncation in the sense that all massive fields interact among themselves 
and with the massless fields.
\subsection{Gauge symmetries of the four-dimensional action.}
Six-dimensional QED has an obvious $U(1)$ symmetry. It is interesting to see how this invariance is traduced in the lower dimensional theory. Before gauge fixing, the four-dimensional action enjoys the infinite set of symmetries:
\bea
&&\d A_\m^n=i\pd_\m \Lambda_n\nonumber\\
&& \d A_5^n=-\frac{n_5}{R_5} \Lambda_n\nonumber\\
&& \d A_6^n =-\frac{n_6}{R_6} \Lambda_n
\eea
Where $\Lambda_n$ are the modes of the expansion of the abelian transformation parameter. All those gauge symmetries $\Lambda_{n_5,n_6}$ are spontaneously broken, except for the
zero mode, corresponding to $\Lambda_{0,0}$. The $A_\m^n$ are the massive vector bosons, and
the $A_5^n$ and $A_6^n$ the scalar higgses.
\par
There is a curious fact, however, and this is the appearance of two singlets in four dimensions,
namely $A_5^0$ and $A_6^0$. Those singlets are massless at tree level, but no symmetry protects
them from getting massive through quantum corrections.
\par
The same fields are protected from getting masses in six dimensions, through gauge invariance
and six dimensional Lorentz covariance. The point is that the breaking
\be
O(1,5)\rightarrow O(1,3)\times O(2)\times O(2)
\ee
of the symmetry group of the vacuum is an instance of spontaneous compactification; i.e., the
equations of motion enjoy the full $O(1,5)$ symmetry, and only the solution breaks it.

\subsection{The massless action}

The zero mode of the above action is
\bea
\lefteqn{S_{zm}=\int d^4 x\left(\bar{\psi}^1\dslash \psi^1+\bar{\psi}^2
\dslash \psi^2+m\left(\bar{\psi}^1\psi^1-\bar{\psi}^2\psi^2\right)
-\frac{1}{2}A_\m\Box A^\m-\right.{}}\nonumber\\
&&\left.-\frac{1}{2}\phi^*\Box\phi-e\left(\bar{\psi}^1\Aslash\psi^1+\bar{\psi}^2\Aslash\psi^2
+\bar{\psi}^1\phi\psi^2-\bar{\psi}^2\phi^*\psi^1\right)\right)
\eea
where we have represented the zero modes of all fields by the same letter without any subindex:
\be
A_5^0-iA_6^0\equiv\phi^0\equiv\phi
\ee
It must be stressed that this is {\em not} a consistent truncation,(in the sense of the word usually employed in supergravity
and superstrings)
owing to the fact that both $A^0_\m$ and $\phi$ couple  diagonally to the whole fermionic tower; it is expected, however, to be a
physically sensible one at energies $E<<M$.
\par
Denoting $\bar{\phi}$ the background for $\phi$ the cuadratic part of the action is
\begin{eqnarray}
\lefteqn{S_{zm}=\int d^4x\left(\bar{\psi}^1\dslash\psi^1+
\bar{\psi}^2\dslash\psi^2+m\left(\bar{\psi}^1\psi^1-
\bar{\psi}^2\psi^2\right)-\frac{1}{2}\phi_\mu\Box\phi^\mu\right.-}{}\nonumber\\
& &{}\left.-\frac{1}{2}\phi^*\Box\phi-e\left(\bar{\psi}^1\bar{\Aslash}\psi^1+
\bar{\psi}^2\bar{\Aslash}\psi^2+\bar{\eta}^1\gamma^\mu\phi_\mu\psi^1+
\bar{\eta}^2\gamma^\mu\phi_\mu\psi^2+\bar{\psi}^1\gamma^\mu\phi_\mu\eta^1+
\right.\right.{}\nonumber\\
& &{}\left.\left.+\bar{\psi}^2\gamma^\mu\phi_\mu\eta^2+\bar{\psi}^1\bar{\phi}\psi^2-
\bar{\psi}^2\bar{\phi}^*\psi^1+\bar{\eta}^1\phi\psi^2-\bar{\eta}^2\phi^*\psi^1+
\bar{\psi}^1\phi\eta^2-\bar{\psi}^2\phi^*\eta^1\right)\right)
\end{eqnarray}
Where $e$ is now the four dimensional coupling. The first two coefficients 
in the heat kernel expansion read:

\be
a_2^{(zm)}=\int\frac{d^4 x}{(4\pi)^2}8\left(m^2 - e^2 |\bar{\phi}|^2\right)
\ee
and
\bea\label{fdcounter}
\lefteqn{a_4^{(zm)}=\int\frac{d^4 x}{(4\pi)^2}\left(\frac{4}{3}e^2\bar{F}_{\mu\nu}^2-
4e^2\bar{\phi}^*\Box\bar{\phi}+8e^2m^2|\bar{\phi}|^2-
4e^4|\bar{\phi}|^4+\right.{}}\nonumber\\
&&+4e^2\left(\bar{\eta}^1\bar{\Dslash}\eta^1+
\bar{\eta}^2\bar{\Dslash}\eta^2\right)+12me^2\left(\bar{\eta}^1\eta^1-\bar{\eta}^2\eta^2\right)+8e^3
\bar{\eta}^2\bar{\phi}^*\eta^1-8e^3\bar{\eta}^1\bar{\phi}\eta^2\Bigg)
\eea
This is the logarithmically divergent  counterterm that arises when renormalizing the zero mode of the four dimensional 
action. 
\par
It should be remarked that  the resulting four dimensional model is superficially 
very similar to the 
Coleman-Weinberg setup, in which
radiative spontaneous symmetry breaking was first discovered. There is a crucial difference
 though, and this is that the scalar field is not charged, in spite of being complex.
The reason is that it remembers its gauge origin, and six-dimensional gauge invariance
manifests here as a Kac-Moody transformation acting on the full tower of 
massive states. In addition to that, the quartic coupling is here a
quantum effect, because it was not present in the bare four-dimensional lagrangian. Also
the scalar field gets massive, with a mass proportional to the fermion mass (times the four-
dimensional fine structure constant)\footnote{At any rate, this yields (twice) the usual beta function for the four dimensional fine structure
 constant:
\be
\b_e=\frac{1}{6\pi^2} e^3
\ee
The behavior of the charge is:
\be
e^2=\frac{e_0^2}{1-\frac{e_0^2}{3\pi^2}\log{\m/\m_0}}
\ee
which blows up at a Landau pole located at
\be
\Lambda\equiv \m_0 e^{3\pi^2 /e_0^2}
\ee
}
.
\section{A comparison of the massless sector of the full six-dimensional divergences 
with the divergences of the massless sector of the four-dimensional theory.}

After all this work, we are finally in a position to study our main concern, namely, how
the divergent part of the six-dimensional effective action is related
to the corresponding four-dimensional quantity.

\subsection{The cutoff theory}
Let us first analyze the problem from the viewpoint of the cutoff theory. As we have seen in six dimensions the divergent part of the effective action is
given through the equation (\ref{6d}); while from the four-dimensional viewpoint the corresponding formula stems from (\ref{4d}).
\par

When we are interested in the zero mode, i.e., the piece in six dimensions where
all fields are independent of the extra dimensions, the measure clearly  factorizes:
\be
d^6 x\rightarrow \frac{1}{M^2}d^4 x
\ee
It is plain that the divergences never coincide exactly.
The {\em only} way to make the divergences related to the fourth heat-kernel coefficient identical
in six and in four dimensions is choose different proper time cutoffs in both dimensions
in such a way that:
\be
\frac{\Lambda_{(d=6)}^2}{M^2}\equiv\log{\frac{\Lambda^2_{(d=4)}}{\m^2_{(d=4)}}}
\ee
We choose that because those coefficients are {\em almost} identical, so that the logarithmic divergences 
are as similar as possible.

This identification leads to the reinterpretation of the six-dimensional quartic divergences as 
$\log^2$:
\be
 \Lambda_{(d=6)}^4\rightarrow M^4 \left(\log{\frac{\Lambda^2_{(d=4)}}{\m^2_{(d=4)}}}\right)^2
\ee
and finally, the six-dimensional logarithmic divergences appear in the guise of log\,log.
\be
\log{\frac{\Lambda^2_{(d=6)}}{\m^2_{(d=6)}}}\rightarrow 
\log\left(\frac{M^2}{\m^2_{(d=6)}}\log{\frac{\Lambda^2_{(d=4)}}{\m^2_{(d=4)}}}\right)
\ee
This reinterpretation gives rise to a few more four-dimensional nonstandard counterterms, 
which we will comment upon in a moment.
\par
Let us stress, for the time being, that the logarithmic divergence, when renormalizing (correctly) from six dimensions
is not identical to  the one (\ref{fdcounter}), but rather
\bea
\lefteqn{\Delta S_{\log}\equiv\int\frac{d^4 x}{(4\pi)^3}e^2
\Bigg(4\left(\bar{\eta}^1\dslash \eta^1+\bar{\eta}^2
\dslash \eta^2 \right)+{}}\nonumber\\
&&+12m\left(\bar{\eta}^1\eta^1-\bar{\eta}^2\eta^2\right)
+\frac{4}{3}\left(\bar{F}^{\m\n}\bar{F}_{\m\n}-\right.\nonumber\\
&&\left.-2\bar{A}_5 \Box \bar{A}_5- 2\bar{A}_6\Box\bar{A}_6\right)-
\nonumber\\
&&-4e\left(\bar{\eta}^1\bar{\Aslash}\eta^1+
\bar{\eta}^2\bar{\Aslash}\eta^2+\bar{\eta}^1\bar{\phi}\eta^2-
\bar{\eta}^2\bar{\phi}^*\eta^1\right)\Bigg)\log{\frac{\Lambda_{d=4}^2}{\m_{d=4}^2}}
\eea
The scalars $A_5$ and $A_6$  are now protected by the six dimensional symmetries, as they should be.

\subsection{Dimensional regularization}
Were we to stick to dimensional regularization, we would have to
 compare the four dimensional counterterm with the massless sector of the 
six-dimensional one, which was previously determined in equation (\ref{sdcounter}). 
There are then 
two types of terms.
\par
First of all, those terms which have negative dimension constants in front, which 
are precisely the ones not present in the original six-dimensional lagrangian, yield in four
dimensions counterterms with dimension six operators, suppressed by two powers
 of the Kaluza-Klein scale:
\begin{eqnarray}
\lefteqn{\Delta S_{(1)}=\frac{e^2}{64\pi^3 M^2 \e}\int d^4 x\left(-\frac{1}{12}e^2\left(\bar{\eta}\Sigma_{\mu\nu\rho}\eta\right)^2+\frac{19}{15}m\bar{\eta}\bar{D}_\m\bar{D}^\m\eta+\right.{}}\nonumber\\
&&\left.+\frac{2}{15}e\bar{\eta}\gamma_\n\bar{D}_\m\eta\bar{F}^{\m\n}-em\bar{\eta}\gamma^\m\gamma^\n\eta\bar{F}_{\m\n}-
\frac{11}{45}\left(\bar{D}_\lambda\bar{F}_{\m\n}\right)^2+
\frac{23}{9}(\bar{D}_\m\bar{F}_{\m\n})^2+\right.\nonumber\\
&&\left.+\ldots\right)
\end{eqnarray}
Where the dots stand for terms with contractions of index in the extra dimensions and $e$ is the four-dimensional coupling. Then, there are the usual four-dimensional counterterms in the guise
\begin{equation}
\Delta S_{(2)}=-\frac{2 e^2 m^2}{64\pi^3 M^2\e}\int d^4 x\left( \bar{\eta}
\bar{\Dslash}\eta+ 3 m\bar{\eta}\eta+\frac{2}{3}
\bar{F}_{\m\n}^2\right)
\end{equation}
The  six-dimensional mass $ m^2$ can clearly be tuned so as to survive in the 
limit in 
which the Kaluza-Klein scale is pushed to infinity. We simply have to tune the dimensionless
quantity 
\be
\frac{e^2 m^2}{64\pi^3 M^2 \e}
\ee
towards the true four-dimensional $\frac{e^2}{16\pi^2 \e}$, while keeping the six-dimensional 
mass $m$
in their four-dimensional value. In such a way we recover {\em almost}
all four dimensional counterterms, albeit with a different sign, which could be accounted for
by changing the direction of the analytical continuation: $\e_{d=6}=-\e_{d=4}$.
\par 
We say almost, because it can easily be seen from these results that there is no room 
for the $|\phi|^2$ and $|\phi|^4$
counterterms, which appear when working upwards from four-dimensions, but do not appear
in the zero mode of the six-dimensional counterterm. 
\par
The only (dim) hope is that these 
four-dimensional counterterms
are actually cancelled when the full tower of Kaluza-Klein states is considered. The next 
subsection is devoted to 
disipate this lingering doubt.
\par
It seems indeed strange that no quartic interaction is generated when coming
from six dimensions. 
No definite conclusions can be draw, however, because those effects are of order $O(\l^2)$,
 where $\l$ is que quartic coupling constant, which means order $O(e^8)$ in our case. We have
no right to keep those terms.
\par
There is a very simple mapping from six-dimensional operators to four-dimensional ones, namely
\be
{\cal O}_{(n)}\rightarrow {\cal O}_{(n-N)}
\ee
where $N$ is the number of fields involved in the operator.
\par
In that way it is seen that the reduction works at follows:
\bea
&&{\cal O}_{(5)}\rightarrow {\cal O}_{(3)}\nonumber\\
&&{\cal O}_{(6)}\rightarrow {\cal O}_{(4)}\nonumber\\
&&{\cal O}_{(7)}\rightarrow {\cal O}_{(5)}\nonumber\\
&&{\cal O}_{(8)}\rightarrow {\cal O}_{(6)}
\eea
except for
\be
{\cal O}^2_{(8)}\rightarrow {\cal O}^2_{(5)}
\ee
In four dimensions, all operators with dimension higher than four appear necessarily
with coefficients which get inverse powers of the compactification scale, $M$. We should be 
then pretty confident of all results gotten in the limit in which this scale goes
to infinity. 
\par
Another question is what happens in the chiral limit. If the mass of the fermion vanishes,
then the six-dimensional counterterms do {\em not} include the four-dimensional ones. If we 
think about it, the conclusion is almost unavoidable, because there is no parameter
in the lagrangian with the dimension of mass. The inverse coupling constant does not 
qualify for this, because it is never going to appear in a perturbative computation.

\section{The full tower of four-dimensional divergences}

Let us consider now the problem of the divergences of the four-dimensional 
 theory with the whole Kaluza-Klein tower. We intend  to compute the counterterm 
asociated with the full four-dimensional Lagrangian (\ref{tower}). 
We let the index $n=(n_5,n_6)$ run over the whole tower of each field. 
Notice that the bosonic fields are now complex (except the one corresponding to $n=0$). 
 $N$ is the complex mass number $N=\frac{n_6}{R_6}+i\frac{n_5}{R_5}$, 
and also   $L=\frac{l_6}{R_6}+i\frac{l_5}{R_5}$.
We have also defined $\bar{\phi}_n\equiv\bar{A}^n_5-i\bar{A}_6^n$ and $\bar{\phi}^*_n\equiv
\bar{A}^n_5+i\bar{A}_6^n\ne(\bar{\phi}_n)^*=\bar{A}^{-n}_5+i\bar{A}_6^{-n}$. 
\par

As we have said the massive ($n\ne0$) modes are complex. In order to use the algorithm explained in the appendix we have to double this modes into real and imaginary parts. However it is also possible to do the calculatios with the complex fields and introduce at the end some extra factors in the adecuate terms. After squaring the matrices and performing the supertraces we get the following 
counterterms in four dimensions with some labor
\be
a_2=\int\frac{d^4 x}{(4\pi)^2}\sum_l\left(8m^2-4|L|^2-8e^2\sum_n\bar{\phi}_n^*\bar{\phi}_{-n}+8e\left(L^*\bar{\phi}_0-L\bar{\phi}^*_0\right)\right)
\ee
The mode sum can be regularized and performed with the help of a zeta function. We shall do it
in the next section, when working out the reduction of $QED_4$ on a two-torus.
The fourth heat-kernel coefficient is \footnote{Here is the explicit expression
\begin{eqnarray}
\lefteqn{a_4=\int\frac{d^4 x}{(4\pi)^2}\sum_l\left(\left(-4m^4+2|L|^4-8m^2|L|^2\right)
+\frac{4}{3}e^2\sum_n\bar{F}_{\mu\nu}^n\bar{F}^{\mu\nu}_{-n}-4e^2\sum_n\bar{\phi}^*_n
\Box\bar{\phi}_{-n}-\right.{}}\nonumber\\
&& \left.-8e\left(m^2+|L|^2\right)\left(L\bar{\phi}_0^*-
L^*\bar{\phi}_0\right)+8e^2\sum_n\left(m^2+|L+N|^2+|N|^2\right)\bar{\phi}^*_n\bar{\phi}_{-n}-\right.\nonumber\\ 
&& \left.-
4e^2\sum_n\left(N^*+L^*\right)L^*\bar{\phi}_n\bar{\phi}_{-n}-4e^2\sum_n\left(N+L\right)L\bar{\phi}_n^*\bar{\phi}_{-n}^*+\right.\nonumber\\
&& +\left.
8e^3\sum_{m,n}\bar{\phi}^*_{m-l}\bar{\phi}_{l-n}\left( M\bar{\phi}^*_{n-m}-
N^*\bar{\phi}_{n-m}\right)+4e^2\sum_{m,n,s}\bar{\phi}_{m-l}^*\bar{\phi}_{l-s}\bar{\phi}^*_{s-n}\bar{\phi}_{n-m}-\right.\nonumber\\
&& \left.-4e^2\sum_nN^*\partial_\mu\bar{\phi}_n\bar{A}_{-n}^\mu+4e^2\sum_nN\partial_\mu
\bar{\phi}_n^*\bar{A}^\mu_{-n}+4e^2\sum_n|N|^2\bar{A}_\mu^n\bar{A}^\mu_{-n}+\right.\nonumber\\
&& \left.+8e^2\sum_{n\ne0}\left(\bar{\eta}^1_{l-n}
\dslash\eta^1_{l-n}+\bar{\eta}^2_{l-n}\dslash\eta^2_{l-n}\right)-
8e^3\sum_{m\ne0,n}\left(\bar{\eta}^1_{l-m}\bar{\Aslash}_{l-n}\eta^1_{n-m}+\bar{\eta}^2_{l-m}\bar{\Aslash}_{l-n}\eta^2_{n-m}\right)+\right.\nonumber\\
&& \left.+
24me^2\sum_{n\ne0}\left(\bar{\eta}^1_{l-n}\eta^1_{l-n}-\bar{\eta}^2_{l-n}\eta^2_{l-n}\right)+
16e^3\sum_{m\ne0,n}\bar{\eta}^2_{l-m}\bar{\phi}^*_{l-n}\eta^1_{n-m}-16e^3\sum_{m\ne0,n}\bar{\eta}^1_{l-m}
\bar{\phi}_{l-n}\eta^2_{n-m}+\right.\nonumber\\&&\left.+16e^2\sum_{n\ne0}L^*\bar{\eta}^2_{l-n}\eta^1_{l-n}+16e^2\sum_{n\ne0}
L\bar{\eta}^1_{l-n}\eta^2_{l-n}+4e^2\left(\bar{\eta}^1_l
\dslash\eta^1_l+\bar{\eta}^2_l\dslash\eta^2_l\right)-\right.\nonumber\\
&& \left.-4e^3\sum_n\left(\bar{\eta}^1_n\bar{\Aslash}_{n-l}\eta^1_l+\bar{\eta}^2_n\bar{\Aslash}_{n-l}\eta^2_l\right)+12me^2\left(\bar{\eta}^1_l\eta^1_l-\bar{\eta}^2_l\eta^2_l\right)+
8e^3\sum_{n}\bar{\eta}^2_n\bar{\phi}^*_{n-l}\eta^1_l-\right.\nonumber\\&&\left.-8e^3\sum_{n}\bar{\eta}^1_n
\bar{\phi}_{n-l}\eta^2_l+8e^2L^*\bar{\eta}^2_l\eta^1_l+8e^2L\bar{\eta}^1_l\eta^2_l
\right)
\end{eqnarray}
}
 quite messy indeed. At least, one thing is clear: there is no way to perform
a clever resummation (like the one Duff and Toms did in the free case) in order  to
 cancel the four dimensional counterterms for both $|\phi|^2$ and $|\phi|^4$, for the simple 
reason that there is no contribution of the massive fields to them. This fact was not
obvious {\em a priori} and the doubt about it was the main reason why 
this computation was performed.

\section{The true four-dimensional renormalization}

From our point of view, in which the full theory is defined in six dimensions, the true
renormalization is the one that is obtained via an harmonic expansion of the 
six-dimensional counterterm(s). 
\subsection{The cutoff theory}
With the interpretation of the six-dimensional cutoff we have advocated, the four-dimensional
logarithmic divergences read 
\bea
\lefteqn{\Delta S_{\log}\equiv\int\frac{d^4 x}{(4\pi)^3}e^2
\sum_{n}\Bigg(4\left(\bar{\eta}^1_{n}\dslash \eta^1_n+\bar{\eta}^2_{n}
\dslash \eta^2_n+N\bar{\eta}^1_{n}\eta^2_n+N^*
\bar{\eta}^2_{n}\eta^1_n\right)+{}}\nonumber\\
&&+12m\left(\bar{\eta}^1_{n}\eta^1_n-\bar{\eta}^2_{n}\eta^2_n\right)
+\frac{4}{3}\left(\bar{F}^{\m\n}_{-n}\bar{F}_{\m\n}^n+2|N|^2\bar{A}_{-n}^\m\bar{A}_\m^n-4i\partial_\m\bar{A}^\m_{-n}\left(\frac{n_5}{R_5}\bar{A}_5^n+\frac{n_6}{R_6}\bar{A}_6^n\right)+\right.\nonumber\\&&\left.
+2\bar{A}_5^{-n}\left(-\Box+\frac{n_6^2}{R_6^2}\right)\bar{A}_5^n+2\bar{A}_6^{-n}\left(-\Box+\frac{n_5^2}{R_5^2}\right)\bar{A}_6^n-4\frac{n_5n_6}{R_5R_6}\bar{A}_5^{-n}\bar{A}_6^n\right)-
\nonumber\\
&&-4e\sum_m\left(\bar{\eta}^1_m\bar{\Aslash}_{m-n}\eta^1_n+
\bar{\eta}^2_m\bar{\Aslash}_{m-n}\eta^2_n+\bar{\eta}^1_m\bar{\phi}_{m-n}\eta^2_n-
\bar{\eta}^2_m\bar{\phi}^*_{m-n}\eta^1_n\right)\Bigg)\log{\frac{\Lambda_{d=4}^2}{\m_{d=4}^2}}
\eea
In addition to that, there are the $\log^2$ divergences, coming from the quartic 
divergences in six dimensions. Those are trivial in our case, because they do not 
depend on the background fields.
\par
Finally, there are the $\log\,\log$ divergences, stemming from the logarithmic divergence
in six dimensions. This divergence is suppressed by the scale of compactification.
The result of a somewhat heavy computation,  keeping terms up to
cubic order in the four-dimensional electric charge, is:
\bea
\lefteqn{\Delta S_{\log\log}\equiv \int\frac{d^4 x}{(4\pi)^3}
\frac{e^2}{M^2}\sum_n\left(-m\bar{\eta}^1_n\left(\frac{19}{15}\left(-\Box+|N|^2\right)+
2m\dslash+6m^2\right)\eta^1_n+\right.{}}\nonumber\\
& &\left.+m\bar{\eta}^2_n\left(\frac{19}{15}\left(-\Box+|N|^2\right)-2m\dslash+6m^2\right)\eta^2_n-2m^2\left(N\bar{\eta}^1_n\eta^2_n+N^*\bar{\eta}^2_n\eta^1_n
\right)+\right.\nonumber\\
&& \left.+\frac{23}{9}\partial_\mu\bar{F}^{\mu\nu}_{-n}\left(\partial^\rho\bar{F}_{\rho\nu}^n-2|N|^2\bar{A}_\n^n\right)-
\bar{F}_{\mu\nu}^{-n}\left(\frac{11}{45}\left(-\Box+|N|^2\right)+\frac{4}{3}m^2\right)\bar{F}^{\mu\nu}_n+\right.\nonumber\\
&&\left.+|N|^2\bar{A}^\mu_{-n}\left(\frac{31}{15}\left(-\Box+
|N|^2\right)-\frac{8}{3}m^2\right)\bar{A}_\mu^n-\right.\nonumber\\
&&\left.-i\partial_\mu\bar{A}^\mu_{-n}\left(\frac{62}{15}\left(
-\Box+|N|^2\right)-\frac{16}{3}m^2\right)\left(\frac{n_5}{R_5}\bar{A}_5^n+\frac{n_6}{R_6}\bar{A}_6^n\right)
+\right.\nonumber\\
&&\left.+\bar{A}_5^{-n}\left(\frac{31}{15}\left(-\Box+|N|^2\right)-\frac{8}{3}m^2\right)\left(-\Box+\frac{n_6^2}{R_6^2}
\right)\bar{A}_5^n+\right.\nonumber\\
&&\left.+\bar{A}_6^{-n}\left(\frac{31}{15}\left(-\Box+|N|^2\right)-\frac{8}{3}m^2\right)\left(-\Box+\frac{n_5^2}{R_5^2}
\right)\bar{A}_6^n-\right.\nonumber\\
&&\left.-\frac{n_5n_6}{R_5R_6}\bar{A}_5^{-n}\left(\frac{62}{15}\left(-\Box+|N|^2\right)-\frac{16}{3}m^2\right)\bar{A}_6^n
\right)\log\left(\frac{M^2}{\m^2_{(d=6)}}\log{\frac{\Lambda^2_{(d=4)}}{\m^2_{(d=4)}}}\right)+\nonumber\\
&&+O(e^3)
\eea

\subsection{Dimensional regularization}
In that case the true divergences only come from the sixth  coefficient,
which yields the $\log\,\log$ divergences we just wrote down.
\par
This means that in addition to the already mentioned counterterms to the zero modes there 
are a full tower of counterterms
involving six-dimensional operators.
\par
It is of  interest to specialize to the massless case ($m=0$), in which, as we have
already noticed, no ordinary dimension four operator is recovered:
\begin{eqnarray}
a_6&=&\int\frac{d^4 x}{(4\pi)^3}
\frac{e^2}{M^2}\sum_n\left(\frac{23}{9}\partial_\mu\bar{F}^{\mu\nu}_{-n}\left(\partial^\rho\bar{F}_{\rho\nu}^n-2|N|^2\bar{A}_\n^n\right)-\frac{11}{45}\bar{F}_{\mu\nu}^{-n}\left(-\Box+|N|^2\right)\bar{F}^{\mu\nu}_n+\right.\nonumber\\
&&\left.+\frac{31}{15}|N|^2\bar{A}^\mu_{-n}\left(-\Box+
|N|^2\right)\bar{A}_\mu^n-i\frac{62}{15}\partial_\mu\bar{A}^\mu_{-n}\left(
-\Box+|N|^2\right)\left(\frac{n_5}{R_5}\bar{A}_5^n+\frac{n_6}{R_6}\bar{A}_6^n\right)
+\right.\nonumber\\
&&\left.+\frac{31}{15}\bar{A}_5^{-n}\left(-\Box+|N|^2\right)\left(-\Box+\frac{n_6^2}{R_6^2}
\right)\bar{A}_5^n+\frac{31}{15}\bar{A}_6^{-n}\left(-\Box+|N|^2\right)\left(-\Box+\frac{n_5^2}{R_5^2}
\right)\bar{A}_6^n-\right.\nonumber\\
&&\left.-\frac{62}{15}\frac{n_5n_6}{R_5R_6}\bar{A}_5^{-n}\left(-\Box+|N|^2\right)\bar{A}_6^n
\right)+O(e^3)
\end{eqnarray}
That is, in the chiral case there is no renormalization of the fermionic tower (at this order) whatsoever,
which is {\em not} what happens from the four-dimensional point of view 
of the previous paragraph.

\section{Another example: $QED_4$ on a two-torus}
Let us now repeat this exercise in a situation that, although
probably much less interesting from the physical point of view, is much better defined as
a quantum theory, namely $QED_4$ on a two-torus. The reduced theory is a two-dimensional one,
where all divergences are more or less trivial (essentially normal ordering). 
It is nevertheless possible to analyze it with the very same general techniques.

\subsection{The four-dimensional viewpoint}

Let us then consider $QED_4$ on a manifold $R^2\times S^1\times S^1$. The action is 
\begin{equation}
S=\int d^4 x\left(\frac{1}{4}F_{\m\n}^2+\bar{\psi}(\Dslash+m)\psi\right)
\end{equation}
where the abelian covariant derivative is simply:
\begin{equation}
D_\m\psi\equiv\left(\pd_\m-eA_\m\right)\psi
\end{equation}
The theory is renormalizable. In dimensional renormalization the counterterm is the fourth
coefficient in the small-time heat kernel expansion: 
\begin{eqnarray}\label{a4}
a_4&=&\int \frac{d^4x}{(4\pi)^{2}}\left(\frac{2}{3}e^2\bar{F}_{\m\n}^2+ 
2e^2\bar{\eta}\gamma^\m\bar{D}_\m\eta+8e^2 m\bar{\eta}\eta\right)\end{eqnarray}
In the cutoff theory, this is precisely the coefficient of the logaritrhmic divergence, but
there is a quadratic divergence as well:
\be
\Delta S=\int d^4 x \left(b_2 \Lambda_{d=4}^2+b_4 
\log\frac{\Lambda_{d=4}^2}{\m_{d=4}^2}\right)
\ee
where
\begin{equation}
a_2=\int \frac{d^4x}{(4\pi)^{2}}4m^2
\end{equation}
\subsection{The two-dimensional viewpoint}
In order to dimensionaly reduce the theory we consider the matrices $(a=1,2)$
\begin{eqnarray}
&&\g_a^{(4)}=\s_3\otimes\s_a\nonumber\\
&&\g_3^{(4)}=\s_1\otimes 1\nonumber\\
&&\g_4^{(4)}=\s_2\otimes1
\eea
In that way, four-dimensional spinors split in two two-dimensional ones:
\be
\psi=\left(\begin{array}{c}
\psi_1\\
\psi_2
\end{array}\right)
\ee

It is a simple matter to perform the integrals over the angular variables and obtain
the gauge fixed action (still exact) in the two-dimensional form:
\bea\label{tower}
\lefteqn{S=\int d^2 x\sum_{n_3,n_4}\left(\bar{\psi}^1_{n}\dslash \psi^1_n+\bar{\psi}^2_{n}
\dslash \psi^2_n+\bar{\psi}^1_{n}(i\frac{n_3}{R_3}+\frac{n_4}{R_4})\psi^2_n-
\bar{\psi}^2_{n}(i\frac{n_3}{R_3}-\frac{n_4}{R_4})\psi^1_n+\right.{}}\nonumber\\
&&{}+m\left(\bar{\psi}^1_{n}\psi^1_n-\bar{\psi}^2_{n}\psi^2_n\right)
-\frac{1}{2}(A_a^n)^{*}\left(\Box-\frac{n_3^2}{R_3^2}-\frac{n_4^2}{R_4^2}\right) A^a_n-
\frac{1}{2}(A_3^n)^{*}\left(\Box-\frac{n_3^2}{R_3^2}-\frac{n_4^2}{R_4^2}\right)A_3^n-
\nonumber\\
&&\left.-\frac{1}{2}(A_4^n)^{*}\left(\Box-\frac{n_3^2}{R_3^2}-
\frac{n_4^2}{R_4^2}\right)A_4^n-e\sum_m\left(\bar{\psi}^1_m\Aslash_{m-n}\psi^1_n+
\bar{\psi}^2_m\Aslash_{m-n}\psi^2_n
+\right.\right.{}\nonumber\\&&{}\left.+\bar{\psi}^1_mA_3^{m-n}\psi^2_n-
\bar{\psi}^2_mA_3^{m-n}\psi^1_n-i\bar{\psi}^1_mA_4^{m-n}\psi^2_n-
i\bar{\psi}^2_mA_4^{m-n}\psi^1_n\right)\Bigg)
\eea
 The two-dimensional coupling constant is
\be
e\equiv \frac{e^{(4)}}{2\pi\sqrt{R_3 R_4}}\equiv e^{(4)} M
\ee
In two dimensions, gauge fields are dimensionless and so are scalar fields. Fermionic fields
enjoy mass dimension $1/2$. We hope that there would arise no confusion 
for the use of the same symbol $e$ for both 
coupling constants.
The zero mode of this action is
\bea
\lefteqn{S=\int d^2 x\left(\bar{\psi}^1\dslash \psi^1+\bar{\psi}^2
\dslash \psi^2+m\left(\bar{\psi}^1\psi^1-\bar{\psi}^2\psi^2\right)
-\frac{1}{2}A_a\Box A^a-\right.{}}\nonumber\\
&&\left.-\frac{1}{2}\phi^*\Box\phi-e\left(\bar{\psi}^1\Aslash\psi^1+\bar{\psi}^2\Aslash\psi^2
+\bar{\psi}^1\phi\psi^2-\bar{\psi}^2\phi^*\psi^1\right)\right)
\eea
where we have represented the zero modes of all fields by the same letter without any subindex:
\be
A_3^0-iA_4^0\equiv\phi^0\equiv\phi
\ee

If we define the theory by dimensional renormalization, the 
 counterterm associated to the above action is 
\be
\Delta S_{zero\,mode}=\frac{1}{\e}a_2^{(0)}=\frac{1}{\e}\int\frac{d^2 x}{4\pi}4
\left(m^2 - e^2 |\phi|^2\right)
\ee

If instead we consider the whole tower the corresponding counterterm is given in terms
of the complex mass parameter:
\be
L\equiv \frac{l_4}{R_4}+i\frac{l_3}{R_3}
\ee
\be
\Delta S_{tower}=\frac{1}{\e}a_2=\frac{1}{\e}\int\frac{d^2 x}{4\pi}
\sum_l4\left(m^2-|L|^2-e^2\sum_n\bar{\phi}_n^*\bar{\phi}_{-n}+e\left(L^*\bar{\phi}_0-
L\bar{\phi}^*_0\right)\right)\ee

Here we have a sum of contributions from all higher modes. This is a divergent sum which needs
regularization. In the expression for the tadpole, for example, we are forced to compute
the sum
\be
T(R)\equiv\sum_{n\in\mathbb{Z}}\frac{n}{R}\equiv \frac{1}{R}\sum_{n\in\mathbb{Z}}n
\ee
This can be regularized, for example, (\cite{Polyakov}) by imposing a cutoff
\bea
\sum_{n=1}n&\equiv& \lim_{\e\rightarrow 0}\sum_{n=1}ne^{-\e n}=\lim_{\e\rightarrow 0}\sum_{n=1}
-\frac{\pd}{\pd \e}e^{-\e n}=-\lim_{\e\rightarrow 0}\frac{\pd}{\pd \e}\sum_{n=1}e^{-\e n}
=\nonumber\\
&=&-\lim_{\e\rightarrow 0}\frac{\pd}{\pd \e}\frac{1}{e^{\e}-1}=
\lim_{\e\rightarrow 0}\frac{e^{\e}}{(e^\e-1)^2}=\lim_{\e\rightarrow 0}\left(\frac{1}{\e^2}-\frac{1}{12}\right)
\eea
This clearly shows the divergence of the sum. When adopting a finite prescription, it is 
important to keep this in mind.
One such finite prescription, quite natural, stems from a consideration of the laplacian
operator 
on the extra dimensions, $\Delta_y$, whose eigenvalues are precisely
\be
\l_l\equiv |L|^2
\ee
and the corresponding $\zeta$ function is
\be
\zeta(s)\equiv \sum_{l\neq 0} \left(|L|^2\right)^{-s}
\ee
which happens to be a particular instance of Epstein's zeta function.
This would lead to definite values for
\be
\sum_l 1\equiv \zeta(s=0)+1=0
\ee

and
\be
\sum_l |L|^2\equiv \zeta(-1)=0
\ee
In order to evaluate the coefficient of the tadpole, it is not possible to use this same
$\zeta$ function. One possibility is to use Riemann's $\zeta$ function
\be
\zeta_R(s)\equiv\sum_{n=1} n^{-s}
\ee
so that, for example,
\be
T(R)=\frac{1}{R}(\zeta_R(-1)-\zeta_R(-1))=0
\ee
Actually this is a unavoidable consequence of any definition in which the first of Hardy's 
properties of the sum of a divergent series is satisfied, namely, if
$\sum a_n=S $ then $\sum \lambda a_n=\lambda S$ (cf.\cite{Hardy}, and the discussion in 
\cite{Meessen})

It has to be acknowledged that the need to use two different zeta functions greatly diminishes
the attractiveness of this whole procedure of resummation.
\par
Ay any rate, in order to eliminate the tadpole, one would have in its case to shift the field:
\be
\bar{\phi}_0\rightarrow\bar{\phi}_0-\frac{T}{2 e}
\ee
This shift would in turn affect the fermionic masses through the Yukawa couplings and 
convey
another contribution to the fermion mass renormalization.
\par 
When either theory is defined through a proper time cutoff, the counterterm is given precisely 
by
\be
\Delta S=a_2\log\frac{\Lambda_{d=2}^2}{\m_{d=2}^2}
\ee

\subsection{The limitations of the two-dimensional approach.}
Let us first concentrate upon dimensional renormalization. The mode expansion of the 
four-dimensional counterterm (\ref{a4}) is:

\bea
\lefteqn{a_4=\int\frac{d^2 x}{(4\pi)^2}\frac{e^2}{M^2}
\sum_{n}\Bigg(2\left(\bar{\eta}^1_{n}\dslash \eta^1_n+\bar{\eta}^2_{n}
\dslash \eta^2_n+N\bar{\eta}^1_{n}\eta^2_n+N^*
\bar{\eta}^2_{n}\eta^1_n\right)+{}}\nonumber\\
&&+8m\left(\bar{\eta}^1_{n}\eta^1_n-\bar{\eta}^2_{n}\eta^2_n\right)
+\frac{2}{3}\left(\bar{F}^{ab}_{-n}\bar{F}_{ab}^n+2|N|^2\bar{A}_{-n}^a\bar{A}_a^n-
4i\partial_a\bar{A}^a_{-n}\left(\frac{n_3}{R_3}\bar{A}_3^n+\frac{n_4}{R_4}\bar{A}_4^n\right)
+\right.\nonumber\\&&\left.
+2\bar{A}_3^{-n}\left(-\Box+\frac{n_4^2}{R_4^2}\right)\bar{A}_3^n+2\bar{A}_4^{-n}\left(-\Box+
\frac{n_3^2}{R_3^2}\right)\bar{A}_4^n-4\frac{n_3n_4}{R_3R_4}\bar{A}_3^{-n}\bar{A}_4^n\right)-
\nonumber\\
&&-2e\sum_m\left(\bar{\eta}^1_m\bar{\Aslash}_{m-n}\eta^1_n+
\bar{\eta}^2_m\bar{\Aslash}_{m-n}\eta^2_n+\bar{\eta}^1_m\bar{\phi}_{m-n}\eta^2_n-
\bar{\eta}^2_m\bar{\phi}^*_{m-n}\eta^1_n\right)\Bigg)
\eea
Which has a zero mode
\bea
a_4^{(0)}&=&\int\frac{d^2 x}{(4\pi)^2}\frac{e^2}{M^2}\Bigg(2\left(\bar{\eta}^1\dslash 
\eta^1+\bar{\eta}^2
\dslash \eta^2\right)+8m\left(\bar{\eta}^1\eta^1-\bar{\eta}^2\eta^2\right)+
\frac{2}{3}\bar{F}^{ab}\bar{F}_{ab}-\frac{4}{3}\bar{\phi}^*\Box\bar{\phi}-\nonumber\\
&&-2e\left(\bar{\eta}^1\bar{\Aslash}\eta^1+\bar{\eta}^2\bar{\Aslash}\eta^2
+\bar{\eta}^1\bar{\phi}\eta^2-\bar{\eta}^2\bar{\phi}^*\eta^1\right)\Bigg)
\eea

In that case, it is plain that there are many differences between the detailed forms of 
the mode expansion
of the renormalized four dimensional theory and the renormalization of the two-dimensional
 mode expansion of the bare four-dimensional theory.

In the cutoff theory we could be tempted to identify
\be
\frac{\Lambda^2_{d=4}}{M^2}\equiv\log\frac{\Lambda^2_{d=2}}{\m_{d=2}^2}
\ee
If one is willing to do this, there are two things that happen. First of all, one never 
recovers the two dimensional correction to the mass of the scalar field,
\be
e^2|\phi|^2
\ee
The reason is exactly the same as it was when reducing from six to four dimensions in our
previous paper, namely, the 
spontaneously nature of the breaking of Lorentz symmetry of the mother theory:
\be
O(1,3)\rightarrow O(1,1)\times O(2)\times O(2)
\ee
It is true that this correction vanishes when one considers the full tower and one is willing
to regularize the sum using the zeta funcion approach. As we have pointed out, there is 
an implicit renormalization of the scalar mass involved in this regularization. It is
nevertheless true that one can regularize the sum in such a way as to get essentially the same
result for the dominant (logarithmic) divergence in both the mother and the daughter theories..
\par
\par
The second thing that happens, and this seems unavoidable, is that  there are 
$\log\,\log\,\Lambda^2$ divergences coming from
the $a_4$ four-dimensional counterterm, suppressed by appropiate powers of the Kaluza-Klein 
scale.
\par 
To conclude, even in this example, 
the two-dimensional theory never forgets its mother. This exercise fully supports the 
general conclusions of our previous reduction.

\section{Conclusions}
Two radically different ways to define $QED_6$ at a one-loop level have been explored. 
The lessons of this exercise seem
to be as follows.
\par
When the fundamental theory is defined
in dimension higher than four using dimensional regularization,
 the divergences of the four dimensional theory do not match
 the ones of the extra-dimensional (mother) one. This is true even in the zero volume limit,
when the volume of the extra dimensions is shrunk to zero, and the Kaluza-Klein scale
correspondingly goes to infinity, and this is also true even when the full Kaluza-Klein
tower is taken into account, as we have shown in detail in an explicit six-dimensional 
example.
\par
In other words, the theory never forgets its higher dimensional origin. This is most clearly
 seen in the chiral limit, but appears also in the massive case, with the need of taking into
account counterterms involving higher dimensional operators, whose coefficients can be computed
in an unambiguous and straightforward way. We understand that a need for those counterterms
has been hinted at in \cite{Oliver} and \cite{Ghilencea}.
\par
The full set of four-dimensional counterterms can be easily recovered from the six-dimensional
one by performing an harmonic expansion. This yields what is, in our opinion, {\em the} correct
way of renormalizing Kaluza-Klein theories.
\par
\par
In the massless case (as well as when coming from an odd number 
of spacetime dimensions) the four-dimensional counterterms are simply not contained in the
higher dimensional ones. The appropiate procedure in those cases would be, from our point of
 view, to compute in the mother theory (in which finite results are obtained through
the use of dimensional regularization), and then perform the mode expansion.
\par
Alternatively, when the quantum theory is defined through a proper time cutoff, we recover the four dimensional 
logarithmic divergences via a tuning of the six-dimensional cutoff. There are then 
calculable $\log\,\log\,\Lambda^2$ divergences coming from the six-dimensional 
logarithmic divergences
as a reminder of the sicknesses of the mother theory. Those divergences are, however, 
suppressed by appropiate powers of the compactification scale, which means that they are multiplied by a small coefficient
at energies at which six-dimensional perturbation theory is reliable (essentially $E/M<<\a_{d=4}^{-1}$).
\par
In neither case do we find from six dimensions corrections to the potential energy
of the four-dimensional singlet scalars associated to the zero modes of the 
extra-dimensional legs of the gauge field. This being true for the zero mode, is clearly a low
energy effect, well within
 the range of validity of the one-loop six-dimensional calculation . 
Those corrections are found in four dimensions because there is no gauge symmetry 
to prevent that to happen.
\par
We have repeated the analysis for $QED_4$ on a two-torus, getting
 similar results. This is very important, because there is now no ambiguity as to how to define the extra-dimensional theory.
 This shows, in our opinion, that our main results do not stem from the ambiguities inherent in any
 practical approach to a non-renormalizable theory.
\par

\par
There are no special difficulties with either odd-dimensional spaces (cf. for example
\cite{Fradkin}) or massless fermions
from the viewpoint of the cutoff theory. Let us finally stress that the strictest equivalence
{\em does work} for free theories coupled to the gravitational field, so that all the 
effects we have studied here are due
to the interaction. 
\par
Our results have obvious applications to the study of the range of validity of
the low energy effective four-dimensional models when studying Kaluza-Klein theories (cf. 
for example 
\cite{Hatanaka})
because our framework is consistent by construction (that is, to the extent that the six-dimensional model is consistent).
\par
Although a very simple abelian model has been studied in this paper 
as an example, we do not expect our main
results to change in more complicated (non abelian) situations.
Besides obvious extensions, like supersymmetry and chiral fermions, it would be interesting 
to study the effects of a nontrivial gravitational field, as well as
the physics of codimension one terms in the action (like the presence of branes in it). 
\par
A most interesting related issue is how the full mother theory compares with the
ultraviolet completion as implied by deconstruction (cf. for example \cite{Pokorski}).
\par
Work
is in progress in several of these matters.
\par
\newpage
\appendix
\section{The one-loop effective action as a superdeterminant.}

In order to get acquainted with the heat kernel techniques  
let us repeat a well-known computation ,
namely, the divergent part of the effective action of quantum electrodynamics in $d=4$ 
dimensions. In doing so, we shall employ a technique first introduced by I. Jack and H. Osborn
 \cite{Jack} (cf. also  \cite{Neufeld}), which is exceedingly convenient in case there are 
non-vanishing fermionic
 background fields (and which is extensively used in the main text). The main idea will 
be to represent the combination of fermionic and
 bosonic determinants as a single superdeterminant, or berezinian, as is sometimes 
refered to.
\par

The  euclidean action reads 
\begin{equation}
\mathcal{L}=\bar{\chi}\gamma^M\partial_M\chi-e\bar{\chi}\gamma^M A_M\chi
+m\bar{\chi}\chi+\frac{1}{4}F_{MN}F^{MN}
\end{equation}
We now split the fields in a classical a a quantum part:
\bea
&&A_{M}=\bar{A}_{M}+\phi_{M}\nonumber\\
&&\chi=\eta+\psi
\eea
where the backgrounds do obey the classical equations of motion, i.e.,
\bea
&&\left(\g^M (\pd_M-e\bar{A}_M)+m\right)\eta=0 \nonumber\\
&&\pd_{M}\bar{F}^{MN}+e \bar{\eta}\g^{N}\eta=0
\eea
Keeping only the terms quadratic in the quantum fields and sticking to Feynman's gauge 
leads to:
\begin{equation}
\mathcal{L}=\bar{\psi}\gamma^M\partial_M\psi-e\bar{\psi}\gamma^M\bar{A}_M\psi+m\bar{\psi}\psi-e\bar{\eta}\gamma^M\phi_M\psi-e\bar{\psi}\gamma^M\phi_M\eta-\frac{1}{2}\phi_M\partial_N\partial^N\phi^M
\end{equation}
which can be written as:
\begin{equation}
\mathcal{L}=\frac{1}{2}\phi_MA_{MN}\phi_N+\bar{\psi}B\psi+\phi_N\bar{\Gamma}_N\psi+\bar{\psi}\Gamma_M\phi_M
\end{equation}
with 
\bea
&&A=-\partial_R\partial^R\delta_{MN}\nonumber\\
&&B=\gamma^M\partial_M-e\gamma^M\bar{A}_M+m\nonumber\\
&&\Gamma_N=-e\gamma_N\eta\hspace{2cm}\nonumber\\
&&\bar{\Gamma}_M=-e\bar{\eta}\gamma_M
\eea
This can equally well be expressed (cf. \cite{Neufeld}) in terms of the supermatrix
\begin{equation}
\Delta=\left( \begin{array}{cc}
A_{MN} & \sqrt{\frac{2}{\mu}}\bar{\Gamma}_M\gamma_5B\gamma_5 \\
\sqrt{2\mu}\Gamma_N & B\gamma_5B\gamma_5 \end{array}\right)
\end{equation}
as
\be
S=\int d^4 x \bar{\xi} \Delta \xi
\ee
with $\xi=(\phi_M,\psi)$.

Our main interest is the computation of the free energy,$ W$:
\be
Z\equiv e^{-W}\equiv e^{-\bar{S}}\int{\cal D}\xi e^{-S[\xi]}
\ee
The free energy 
\be
W\equiv\log\, Z
\ee
is then given by
\be
W\equiv \bar{S}+\frac{1}{2}\log\,sdet\,\Delta
\ee 
The superdeterminant, or berezinian of a supermatrix $M$ 
involving bosonic ($+$) and fermionic ($-$)
entries
\bea
M=
\left(
\begin{array}{cc}
M_{++}&M_{+-}\\
M_{-+}&M_{--}
\end{array}\right)
\eea
is defined by
\be
ber\,M\equiv sdet\,M\equiv det\,M_{++}det^{-1}\left(M_{--}-M_{-+}M_{++}^{-1}M_{+-}\right)
\ee
We have introduced an arbitrary mass scale $\m$ for dimensional reasons.
In the present situation,
\begin{equation}\label{sdet}
\Delta=\left( \begin{array}{cc}
-\partial_R\partial^R\delta_{MN} &  \sqrt{\frac{2}{\mu}}e\left( \bar{\eta}\gamma^M\gamma^R\bar{D}_R-m\bar{\eta}\gamma^M\right) \\
-\sqrt{2\mu}e\gamma_N\eta & -\bar{D}_M\bar{D}^M+\frac{e}{2}\gamma^M\gamma^N\bar{F}_{MN}+m^2 \end{array}\right)
\end{equation}
This supermatrix operator enjoys a  Laplacian  form $\Delta=-D_MD^M+Y$ with $D_M=\partial_M+X_M$ and the supermatrices
\begin{equation}
X_M=\left( \begin{array}{cc}
0 & \frac{-e}{\sqrt{2\mu}}\bar{\eta}\gamma^R\gamma_M \\
0 & -e\bar{A}_M \end{array}\right)
\end{equation}
And
\begin{equation}
Y=\left( \begin{array}{cc}  
0 & \frac{-e}{\sqrt{2\mu}}\left( 2m\bar{\eta}\gamma^M+\bar{D}_R\bar{\eta}\gamma^M\gamma^R\right) \\
-\sqrt{2\mu}e\gamma_N\eta & \frac{e}{2}\gamma^M\gamma^N\bar{F}_{MN}+m^2 \end{array}\right)
\end{equation}
Once we have reduced our problem to the computation of the determinant of a
supermatrix the divergent part of the effective action is given by the $a_4(\Delta)$ coefficient in the heat kernel expansion
\begin{equation}
a_4(\Delta)=\int \frac{d^dx}{(4\pi)^{d/2}}\,str\left(\frac{1}{2}Y^2+\frac{1}{12}X_{MN}^2\right)
\end{equation}
Where as usual $X_{MN}$ is the field strength asociated with $X_M$ and $str$ denotes super trace. 
In our case the field strength supermatrix is
\begin{equation}
X_{MN}=\left( \begin{array}{cc}
0 & \frac{-e}{\sqrt{2\mu}}\left( \bar{D}_M\bar{\eta}\gamma^R\gamma_N-\bar{D}_N\bar{\eta}\gamma^R\gamma_M\right) \\
0 & -e\bar{F}_{MN} \end{array}\right)
\end{equation}
Which after squaring and tracing gives a contribution
\begin{equation}
\frac{1}{12}str\,X_{MN}^2=-\frac{2^{[d/2]}}{12}e^2\bar{F}_{MN}^2
\end{equation}
While the contribution from $Y^2$ is
\begin{equation}
\frac{1}{2}str\,Y^2=e^2(d-2)\bar{\eta}\gamma^M\partial_M\eta-e^3 (d-2)\bar{\eta}\gamma^M\bar{A}_M\eta+2de^2m\bar{\eta}\eta+\frac{2^{[d/2]}}{4}e^2\bar{F}_{MN}^2
\end{equation}
Finally we can write the coefficient in four dimensions
\begin{eqnarray}
a_4(\Delta)&=&\int \frac{d^4x}{(4\pi)^{2}}\left(\frac{2}{3}e^2\bar{F}_{MN}^2+2e^2\bar{\eta}\gamma^M\bar{D}_M\eta+8e^2 m\bar{\eta}\eta\right)
\end{eqnarray}
which coincides with the result obtained through the application of the classical 't
 Hooft algorithm \cite{Omote},\cite{Hooft}.

\section*{Acknowledgments}

This work has been partially supported by the
European Commission (HPRN-CT-200-00148) and by FPA2003-04597 (DGI del MCyT, Spain).      
A.F. Faedo has been supported by a MEC grant, AP-2004-0921. We are indebted to Jorge Conde, 
C\'esar G\'omez and Karl Landsteiner
 for useful discussions.


\end{document}